\documentclass[12pt,aps,showpacs,notitlepage]{revtex4-1}
\usepackage{amssymb,amsmath}
\newcommand{\beq}{\begin{equation}}
\newcommand{\beqn}{\begin{eqnarray}}
\newcommand{\eeq}{\end{equation}}
\newcommand{\eeqn}{\end{eqnarray}}
\begin{document}

\title{Conformal Transformations with Multiple Scalar Fields}
\author{David I. Kaiser}
\email{Email: dikaiser@mit.edu}
\affiliation{Center for Theoretical Physics and Department of Physics, \\
Massachusetts Institute of Technology, Cambridge, Massachusetts 02139 USA}
\date{\today}

\begin{abstract} 
Many interesting models incorporate scalar fields with non-minimal couplings to the spacetime Ricci curvature scalar. As is well known, if only one scalar field is non-minimally coupled, then one may perform a conformal transformation to a new frame in which both the gravitational portion of the Lagrangian and the kinetic term for the (rescaled) field assume canonical form. We examine under what conditions the gravitational and kinetic terms in the Lagrangian may be brought into canonical form when more than one scalar field has non-minimal coupling. A particular class of two-field models admits such a transformation, but models with more than two non-minimally coupled fields in general do not.
\end{abstract}
\pacs{04.50.-h; 04.62.+v; 98.80.-k. Preprint MIT-CTP 4125. \\ Published as Phys. Rev. D81, 084044 (2010). }
\maketitle

\section{Introduction} 

Scalar fields with non-minimal couplings to the spacetime Ricci curvature scalar are ubiquitous in particle physics and cosmology. Such non-minimal couplings are fairly generic \cite{FujiiMaeda,Faraoni2004}: they appear in scalar-tensor theories such as Jordan-Brans-Dicke gravity \cite{BransDicke} and induced-gravity models \cite{InducedGravity}; in the low-energy effective actions arising from higher-dimensional theories such as supergravity, string theory, and other Kaluza-Klein models \cite{KaluzaKlein,Zhuk}; and in $f(R)$ models of gravity following a conformal transformation \cite{Sotiriou}. More generally, as has been established for some time, non-minimal couplings necessarily arise as counter-terms when considering the renormalization of scalar fields in curved background spacetimes \cite{BirrellDavies,Buchbinder}. Indeed, in many models the non-minimal coupling strength, $\xi$, grows without bound under renormalization-group flow \cite{Buchbinder}.

Many models have been studied of cosmic inflation driven by a non-minimally coupled scalar field \cite{NMCinfl}, including extended inflation \cite{extendedinfl} and induced-gravity inflation \cite{IgI}. (For reviews, see \cite{Faraoni2001,Faraoni2004}.) Recent work suggests the exciting possibility that the Higgs sector from the electroweak Standard Model could support a viable early-universe phase of inflation as well, provided the Higgs sector is non-minimally coupled \cite{ShapBezrukov}. This recent model has been dubbed ``Higgs inflation" \cite{HiggsInflation,Burgess,Hertzberg}.

Nearly all of the analyses of ``Higgs inflation" have tacitly adopted the unitary gauge, in which only the (real) Higgs scalar field survives and no Goldstone fields remain in the spectrum. The model then reduces to a single-field  case, akin to those reviewed in \cite{Faraoni2004,Faraoni2001}. One may perform a familiar conformal transformation on the spacetime metric, $g_{\mu\nu} \rightarrow \hat{g}_{\mu\nu}$, to bring the gravitational portion of the Lagrangian into the Einstein-Hilbert form. One may also rescale the scalar field, $\phi \rightarrow \hat{\phi}$, so that the kinetic term for $\hat{\phi}$ in the transformed Lagrangian appears in canonical form. Then the system in the transformed frame behaves just like a minimally coupled scalar field in ordinary (Einsteinian) gravity \cite{BirrellDavies,Maeda89,Faraoni98,FujiiMaeda,Faraoni2004,Shapiro}.

Yet the unitary gauge is not renormalizable, and thus it is inappropriate for studies of Higgs-sector dynamics far above the symmetry-breaking scale. To study inflationary dynamics in ``Higgs inflation," one must instead use a renormalizable gauge, in which the Goldstone scalar fields remain explicit \cite{WeinbergQFT}. We are forced, in other words, to consider a multi-field model involving four real scalar fields (the Higgs scalar plus three Goldstone scalars), each of which is non-minimally coupled to the Ricci curvature scalar. As recently noted \cite{Burgess,Hertzberg}, for the model of ``Higgs inflation," no combination of conformal transformation and rescaling of the scalar fields exists that could bring both the gravitational portion of the Lagrangian and the kinetic terms for each scalar field into canonical form.

Building on this important observation, we consider under what conditions a combination of conformal transformation and field rescalings could bring both the gravitational and kinetic terms of a Lagrangian into canonical form, for arbitrary numbers of non-minimally coupled scalar fields. (See also \cite{DamourNordvedt92} on post-Newtonian parameters for tensor-multiscalar models.) Because non-minimal couplings are generic for scalar fields in curved spacetimes --- and because realistic models of particle physics (including generalizations of the Standard Model) contain many scalar fields that could play important roles in the early universe \cite{InflParticlePhys} --- it is important to understand the transformation properties of arbitrary models. 

As we will see, only a particular class of models involving two non-minimally coupled scalar fields admits the desired transformation; models involving more than two non-minimally coupled scalar fields in general do not. One may of course always perform a conformal transformation on the spacetime metric to work in a convenient frame. What one cannot do, in general, is find such a transformed frame in which both the gravitational sector and the scalar fields' kinetic terms assume canonical form.

In Section II we consider the single-field case, to review the usual transformation and clarify notation. In Section III we consider $N$ scalar fields with non-minimal couplings, distinguishing between the cases of $N = 2$ and $N > 2$. Conclusions follow in Section IV.

\section{Single-Field Case}

We will work in $D$ spacetime dimensions (only one of which is timelike); our metric has signature $(-,+,+,+,+,...)$. We take the Christoffel symbols to be
%%%%%%%%%%%%%%
\beq
\Gamma^\lambda_{\mu\nu} = \frac{1}{2} g^{\lambda \sigma} \left[ \partial_\mu g_{\sigma \nu} + \partial_\nu g_{\sigma \mu} - \partial_\sigma g_{\mu\nu} \right] ,
\label{Christoffel}
\eeq
and the Riemann tensor to be
%%%%%%%%%%%%%%
\beq
R^\lambda_{\>\>\> \mu\nu\sigma} = \partial_\nu \Gamma^\lambda_{ \mu\sigma} - \partial_\sigma \Gamma^\lambda_{\mu\nu} + \Gamma^\eta_{\mu \sigma} \Gamma^\lambda_{\eta \nu} - \Gamma^\eta_{ \mu\nu} \Gamma^\lambda_{ \eta \sigma} .
\label{Riemanndef}
\eeq
The Ricci tensor and Ricci curvature scalar follow upon contractions of the Riemann tensor:
%%%%%%%%%
\beq
\begin{split}
R_{\mu\nu} &= R^\lambda_{\>\>\> \mu\lambda \nu} , \\
R &= g^{\mu\nu} R_{\mu\nu} .
\end{split}
\label{Riccidef}
\eeq

In the single-field case the action is given by
%%%%%%%%%%%
\beq
S = \int d^D x \sqrt{-g} \left[ f(\phi) R - \frac{1}{2} g^{\mu\nu} \nabla_\mu \phi \nabla_\nu \phi - V (\phi) \right] .
\label{action}
\eeq
Covariant derivatives are denoted by $\nabla$. We will assume that $f (\phi)$ is positive definite. Minimal coupling corresponds to $f (\phi) \rightarrow (16 \pi G_D )^{-1}$, where $G_D$ is the value of the gravitational constant (akin to Newton's constant) in $D$ dimensions. The frame in which $f (\phi) \neq {\rm constant}$ appears in the action, as in Eq. (\ref{action}), is often referred to as the Jordan frame.

We will assume natural units ($c = \hbar = 1$) and take the metric tensor, $g_{\mu\nu}$, to be dimensionless. Then in $D$ dimensions, times and lengths have dimensions of $(mass)^{-1}$, and the covariant volume element in the action integral, $d^D x \sqrt{-g}$, assumes dimensions of $(mass)^{-D}$. The Ricci scalar, $R (g_{\mu\nu})$, has dimensions $[(\partial_x g )^2 ] \sim (mass)^2$. In order for the integrand to remain dimensionless, meanwhile, the kinetic term for the scalar field requires that $\phi$ have dimensions $[ \phi ] \sim (mass)^{(D - 2)/2}$. We may further parameterize
%%%%%%%%%
\beq
M_{(D)}^{D-2} \equiv \frac{1}{8\pi G_D} ,
\label{MD}
\eeq
in terms of a (reduced) Planck mass in $D$ dimensions. When $D = 4$, we have $M_{(4)} = M_{\rm pl} = 1/\sqrt{8\pi G_4} = 2.43 \times 10^{18}$ GeV. 

Many families of models have an action in the form of Eq. (\ref{action}), in which the scalar field enters with canonical kinetic term but the gravitational sector departs from Einstein-Hilbert form. For example, the non-minimal coupling associated with the renormalization counter-term takes the form
%%%%%%%%%%%%
\beq
f (\phi) = \frac{1}{2} \left[ M_0^{D-2} + \xi \phi^2 \right] ,
\label{fsingle}
\eeq
where $\xi$ is the non-minimal coupling strength and $M_0$ is some mass scale. In the sign conventions of Eq. (\ref{fsingle}), a conformally coupled field has $\xi = - \frac{1}{4} (D - 2) / (D - 1)$. The mass scale $M_0$ need not be identical with $M_{(D)}$. If the scalar field's potential, $V(\phi)$, admits symmetry-breaking solutions with some non-zero vacuum expectation value, $v$, then the measured strength of gravity following symmetry-breaking would be $M_{(D)}^{D-2} = M_0^{D-2} + \xi v^2$. One could even have $M_0 = 0$, as in induced-gravity models \cite{InducedGravity,IgI}.

Another common model is Jordan-Brans-Dicke gravity \cite{BransDicke}, the action for which is often written as
%%%%%%%%%%
\beq
S_{JBD} = \int d^D x \sqrt{-g} \left[ \Phi R - \frac{\omega}{\Phi} g^{\mu\nu} \nabla_\mu \Phi \nabla_\nu \Phi \right] .
\label{SJBD}
\eeq
Rescaling the field, $\Phi \rightarrow \phi^2 / (8 \omega)$, puts the action in the form of Eq. (\ref{action}), with canonical kinetic term for $\phi$ and non-minimal coupling, $f (\phi)$, as in Eq. (\ref{fsingle}), with $M_0 = 0$ and $\xi = 1/(8\omega)$.

We may make a conformal transformation of the metric, defined as
%%%%%%%%%
\beq
\hat{g}_{\mu\nu} = \Omega^2 (x) g_{\mu\nu} .
\label{conftrans}
\eeq
We assume that $\Omega (x)$ is real and therefore that $\Omega^2 (x)$ is positive definite. Note that we have not made a coordinate transformation; the coordinates $x^\mu$ remain the same in each frame. We have instead mapped one metric into another, in a manner that depends on space and time \cite{Wald}. We will use a caret to indicate quantities in the transformed frame. From Eq. (\ref{conftrans}), we immediately see that
%%%%%%%%%%%
\beq
\begin{split}
\hat{g}^{\mu\nu} &= \frac{1}{\Omega^2 (x)} g^{\mu\nu} , \\
\sqrt{-\hat{g}} &= \Omega^D (x) \sqrt{-g} .
\end{split}
\label{raiselower}
\eeq

Upon making the transformation of Eq. (\ref{conftrans}), one may compute the Christoffel symbols and the Ricci curvature scalar in the new frame. One finds \cite{BirrellDavies,Wald,Faraoni98,FujiiMaeda,Faraoni2004} 
%%%%%%%%%%%%
\beq
\begin{split}
\hat{\Gamma}^\alpha_{\beta\gamma} &= \Gamma^\alpha_{\beta \gamma} + \frac{1}{\Omega} \left[ \delta^\alpha_\beta \nabla_\gamma \Omega + \delta^\alpha_\gamma \nabla_\beta \Omega - g_{\beta\gamma} \nabla^\alpha \Omega \right] , \\
\hat{R} &= \frac{1}{\Omega^2} \left[ R -  \frac{2 (D - 1)}{\Omega} \Box \Omega  - \left( D - 1 \right) \left( D - 4 \right) \frac{1}{\Omega^2} g^{\mu\nu} \nabla_\mu \Omega \nabla_\nu \Omega \right] ,
\end{split}
\label{hatR}
\eeq
where
%%%%%%%%%%%
\beq
\Box \Omega = g^{\mu\nu} \nabla_\mu \nabla_\nu \Omega = \frac{1}{\sqrt{-g}} \partial_\mu \left[ \sqrt{-g} g^{\mu\nu} \partial_\nu \Omega \right] .
\label{box}
\eeq
One must be careful to specify whether one is taking derivatives with respect to the original metric, $g_{\mu\nu}$, or the transformed metric, $\hat{g}_{\mu\nu}$, because the Christoffel symbols (and hence covariant derivatives) transform in $\Omega$-dependent ways under the transformation of Eq. (\ref{conftrans}). 

Using Eqs. (\ref{conftrans}) - (\ref{box}), we may rewrite the first term in the action, involving $R$: 
%%%%%%%%%%%%
\beq
\int d^D x \sqrt{-g} f (\phi) R = \int d^D x \frac{\sqrt{-\hat{g}}}{\Omega^D} f (\phi)  \left[ \Omega^2 \hat{R} + \frac{2 \left( D - 1\right) }{\Omega} \Box \Omega + \frac{\left( D - 1\right) \left( D - 4 \right) }{\Omega^2} g^{\mu\nu} \nabla_\mu \Omega \nabla_\nu \Omega \right] .
\label{action1}
\eeq
Let us look at each of these terms in turn. The first term on the righthand side becomes
%%%%%%%%%%
\beq
\int d^D x \sqrt{-\hat{g}} \left[ \left( \frac{f}{\Omega^{D - 2}} \right) \hat{R} \right] .
\eeq
To obtain the canonical Einstein-Hilbert gravitational action in the transformed frame, we identify
%%%%%%%%%%%
\beq
\Omega^{D - 2} (x) = \frac{2}{M_{(D)}^{D-2}} f [\phi (x)] .
\label{Omega}
\eeq

We may integrate the second term on the righthand side of Eq. (\ref{action1}) by parts. Note that the $\Box$ operator acting on $\Omega$ is defined in terms of the original metric, $g_{\mu\nu}$, rather than the transformed metric. Using Eqs. (\ref{raiselower}), (\ref{box}), and (\ref{Omega}), we find
%%%%%%%%%%
\beq
\begin{split}
&\int d^D x \sqrt{-\hat{g}} \> \frac{2 \left( D - 1 \right) }{\Omega^{D+1}} f \> \Box \Omega \\
&\quad\quad = - \int d^D x \sqrt{-\hat{g}} \> (D - 1) (D - 3) M_{(D)}^{D-2} \frac{1}{\Omega^2} \hat{g}^{\mu\nu} \hat{\nabla}_\mu \Omega \hat{\nabla}_\nu \Omega .
\end{split}
\label{action2}
\eeq
Recall that $x^\mu$ is unaffected by the conformal transformation, so that $\hat{\partial}_\mu = \partial_\mu$. Because the covariant derivatives in Eq. (\ref{action2}) act only on scalar functions, we have $\nabla_\mu \Omega = \partial_\mu \Omega$, and hence $\hat{\nabla}_\mu \Omega = \nabla_\mu \Omega$.

The last term on the righthand side of Eq. (\ref{action1}) is
%%%%%%%%%%
\beq
\begin{split}
&\int d^D x \sqrt{-\hat{g}} \> (D - 1) (D - 4) \left( \frac{f}{\Omega^{D+2}} \right) g^{\mu\nu} \nabla_\mu \Omega \nabla_\nu \Omega \\
&\quad\quad  = \int d^D x \sqrt{-\hat{g}} \> \frac{1}{2} (D - 1 ) (D - 4) M_{(D)}^{D-2} \> \frac{1}{\Omega^2} \> \hat{g}^{\mu\nu} \hat{\nabla}_\mu \Omega \hat{\nabla}_\nu \Omega ,
\end{split}
\label{action3}
\eeq
where we have again used Eqs. (\ref{raiselower}) and (\ref{Omega}). Combining Eqs. (\ref{action1}), (\ref{action2}), and (\ref{action3}) (and noting a simple algebraic relationship among the coefficients in front of the $\hat{\nabla}_\mu \Omega$ terms), we find 
%%%%%%%%%%%
\beq
\begin{split}
\int d^D x \sqrt{-g} f (\phi)  R &= \int d^D x \sqrt{-\hat{g}} \> \frac{M_{(D)}^{D-2}}{2} \left[  \hat{R} -  (D - 1) (D - 2) \frac{1}{\Omega^2} \hat{g}^{\mu\nu} \hat{\nabla}_\mu \Omega \hat{\nabla}_\nu \Omega \right] .
\end{split}
\label{actionRhat}
\eeq
The gravitational portion of the action now includes a canonical Einstein-Hilbert term. For this reason, the frame corresponding to $\hat{g}_{\mu\nu}$ is often referred to as the Einstein frame.

We may next consider how the scalar field's kinetic and potential terms in the action transform under $g_{\mu\nu} \rightarrow \hat{g}_{\mu\nu}$. We have
%%%%%%%%%%%
\beq
\begin{split}
\int d^D x \sqrt{-g} & \left[ - \frac{1}{2} g^{\mu\nu} \nabla_\mu \phi \nabla_\nu \phi - V (\phi) \right] 
= \int d^D x \sqrt{-\hat{g}} \left[ - \frac{1}{2} \frac{1}{\Omega^{D-2}} \hat{g}^{\mu\nu} \hat{\nabla}_\mu \phi \hat{\nabla}_\nu \phi - \hat{V} \right] ,
\end{split}
\label{actionmatter}
\eeq
where we have introduced a transformed potential,
%%%%%%%%%%%
\beq
\hat{V} \equiv \frac{V}{\Omega^D} .
\label{Vhat}
\eeq

The full action of Eq. (\ref{action}) may then be written
%%%%%%%%%%%%
\beq
\begin{split}
\int d^D x & \sqrt{-g} \left[ f (\phi)  R - \frac{1}{2} g^{\mu\nu} \nabla_\mu \phi \nabla_\nu \phi - V \right] 
\\
\quad\quad &= \int d^D x \sqrt{-\hat{g}} \left[ \frac{M_{(D)}^{D-2}}{2} \hat{R} - \frac{1}{2} (D - 1) (D - 2) M_{(D)}^{D-2} \frac{1}{\Omega^2} \hat{g}^{\mu\nu} \hat{\nabla}_\mu \Omega \hat{\nabla}_\nu \Omega \right. \\
&\quad\quad\quad\quad\quad\quad\quad\quad  \left. - \frac{1}{2} \frac{1}{\Omega^{D-2}} \hat{g}^{\mu\nu} \hat{\nabla}_\mu \phi \hat{\nabla}_\nu \phi - \hat{V} \right] .
\end{split}
\label{actionsingle1}
\eeq
Upon substituting $f$ for $\Omega$ using Eq. (\ref{Omega}), the action in the transformed frame becomes
%%%%%%%%%
\beq
\int d^D x \sqrt{-\hat{g}}  \left[ \frac{M_{(D)}^{D-2}}{2}  \hat{R} -  \frac{1}{2} \frac{(D-1)}{(D-2)}  M_{(D)}^{D-2} \frac{1}{f^2} \hat{g}^{\mu\nu} \hat{\nabla}_\mu f \hat{\nabla}_\nu f - \frac{1}{4f} M_{(D)}^{D-2}  \>  \hat{g}^{\mu\nu} \hat{\nabla}_\mu \phi \hat{\nabla}_\nu \phi  - \hat{V} \right] .
\label{actionsingle2}
\eeq

In the single-field case, we may next rescale the field, $\phi \rightarrow \hat{\phi}$, so that the new scalar field in the transformed frame has the canonical kinetic term in the action of Eq. (\ref{actionsingle2}). We define $\hat{\phi}$ such that
%%%%%%%%%%
\beq
-\frac{1}{2} \hat{g}^{\mu\nu} \hat{\nabla}_\mu \hat{\phi} \hat{\nabla}_\nu \hat{\phi} = - \frac{M_{(D)}^{D-2}}{4f} \hat{g}^{\mu\nu} \left[ \hat{\nabla}_\mu \phi \hat{\nabla}_\nu \phi + \frac{2 (D - 1)}{(D - 2)}\frac{1}{f} \hat{\nabla}_\mu f \hat{\nabla}_\nu f \right] .
\label{phihat1}
\eeq
In the single-field case, we may assume a one-to-one mapping between $\hat{\phi}$ and $\phi$; in particular, we assume that $\hat{\phi} \rightarrow \hat{\phi} (\phi)$, or
%%%%%%%%%%
\beq
\frac{d\hat{\phi}}{d\phi} = F (\phi) ,
\eeq
in terms of some as-yet unspecified function $F$. In the single-field case, we also have $f = f(\phi)$, so that Eq. (\ref{phihat1}) yields 
%%%%%%%%%%
\beq
F (\phi) = \left( \frac{d\hat{\phi}}{d\phi} \right) =  \sqrt{\frac{M_{(D)}^{D-2}}{2 f^2 (\phi )} } \sqrt{ f (\phi) + \frac{2 (D -1)}{(D - 2)} \left[ f' (\phi ) \right]^2} ,
\label{phihat3}
\eeq
where primes denote derivatives with respect to $\phi$. In terms of the rescaled field, the action of Eq. (\ref{actionsingle2}) may be written
%%%%%%%%%%%
\beq
\begin{split}
\int d^D x \sqrt{-g} &\left[ f(\phi) R - \frac{1}{2} g^{\mu\nu} \nabla_\mu \phi \nabla_\nu \phi - V (\phi) \right] \\
&= \int d^D x \sqrt{-\hat{g}} \left[ \frac{M_{(D)}^{D-2}}{2} \hat{R} - \frac{1}{2} \hat{g}^{\mu\nu} \hat{\nabla}_\mu \hat{\phi} \hat{\nabla}_\nu \hat{\phi} - \hat{V} (\hat{\phi} ) \right] .
\end{split}
\label{actionsingleEinstein}
\eeq
The action in the second line now has both the canonical Einstein-Hilbert form for the gravitational portion as well as the canonical kinetic term for the scalar field. 

Because we are only considering models in which $f (\phi )$ is positive definite and real, the combination on the righthand side of Eq. (\ref{phihat3}) is always non-zero. Under these conditions, models with a single non-minimally coupled scalar field may be related, via conformal transformation and field rescaling, to an equivalent model involving ordinary gravity and a minimally coupled scalar field. (If one relaxes the conditions on $f(\phi)$, and/or considers models with non-canonical kinetic terms in the Jordan frame, then one may find models for which $F (\phi)$ in Eq. (\ref{phihat3}) vanishes. Following a conformal transformation, such models are equivalent to Einstein gravity with a cosmological constant. \cite{Shapiro})

\section{Multi-Field Case}

Let us now consider the case of multiple scalar fields, each with its own non-minimal coupling to $R$. We will use Latin indices to label directions in field space:  $\phi^i$, with $i = 1, ... , N$. Working again in $D$ spacetime dimensions, the action in the Jordan frame takes the form
%%%%%%%%%%%%%%
\beq
\int d^D x \sqrt{-g} \left[ f (\phi^1,  ... , \phi^N ) R - \frac{1}{2} \delta_{ij} g^{\mu\nu} \nabla_\mu \phi^i \nabla_\nu \phi^j - V (\phi^1,  ..., \phi^N ) \right] .
\label{actionmultiJordan}
\eeq
Just as in Eq. (\ref{conftrans}), we may make a conformal transformation, $g_{\mu\nu} \rightarrow \hat{g}_{\mu\nu}$ in terms of some conformal factor, $\Omega^2 (x)$. All of the steps that led from Eq. (\ref{raiselower}) to Eq. (\ref{action3}) depended only on the fact that the terms $\Omega (x)$ and $f(x)$ depended on $x^\mu$; we did not use their functional dependence on the scalar field, $\phi (x)$. Those steps therefore proceed in precisely the same way in the multi-field case, and we again arrive at the gravitational portion of the action in the new frame: 
%%%%%%%%%%%
\beq
\int d^D x \sqrt{-g} \> f (\phi^i ) R = \int d^D x \sqrt{-\hat{g}} \left[ \frac{M_{(D)}^{D-2}}{2} \hat{R} - \frac{1}{2} \frac{(D - 1)}{(D - 2)} M_{(D)}^{D-2} \frac{1}{f^2} \hat{g}^{\mu\nu} \hat{\nabla}_\mu f \hat{\nabla}_\nu f \right] ,
\label{Rhat2}
\eeq
upon using Eq. (\ref{Omega}) to substitute $f$ for $\Omega$. The kinetic and potential terms for the scalar fields transform similarly to Eq. (\ref{actionmatter}), and we find
%%%%%%%%%%%%
\beq
\int d^D x \sqrt{-g} \left[ - \frac{1}{2} \delta_{ij} g^{\mu\nu} \nabla_\mu \phi^i \nabla_\nu \phi^j - V (\phi^i ) \right] = \int d^D x \sqrt{-\hat{g}} \left[ - \frac{1}{4f} M_{(D)}^{D-2} \delta_{ij} \hat{g}^{\mu\nu} \hat{\nabla}_\mu \phi^i \hat{\nabla}_\nu \phi^j - \hat{V} \right] ,
\label{mattermulti}
\eeq
in terms of $\hat{V}$ as defined in Eq. (\ref{Vhat}).

Combining terms, we find the action in the transformed frame
%%%%%%%%%%%%
\beq
\begin{split}
&\int d^D x \sqrt{-g} \left[ f (\phi^i) R - \frac{1}{2} \delta_{ij} g^{\mu\nu} \nabla_\mu \phi^i \nabla_\nu \phi^j - V \right] \\
&\quad = \int d^D x \sqrt{-\hat{g}} \left[ \frac{M_{(D)}^{D-2}}{2} \hat{R} - \frac{1}{2} \frac{(D - 1)}{(D - 2)}\frac{ M_{(D)}^{D-2} }{f^2} \hat{g}^{\mu\nu} \hat{\nabla}_\mu f \hat{\nabla}_\nu f - \frac{M_{(D)}^{D-2}}{4f} \delta_{ij} \hat{g}^{\mu\nu} \hat{\nabla}_\mu \phi^i \hat{\nabla}_\nu \phi^j - \hat{V} \right] .
\end{split}
\label{actionmulti1}
\eeq
In the multi-field case we have $f = f (\phi^1, ... , \phi^N)$, and thus
%%%%%%%%%
\beq
\hat{\nabla}_\mu f =  \left( \hat{\nabla}_\mu \phi^i \right) f_{,i} ,
\label{nablafmulti}
\eeq
where $f_{,i} = \partial f / \partial \phi^i$. We may therefore rewrite the derivative terms in the bottom line of Eq. (\ref{actionmulti1}) in terms of a metric in field space, ${\cal G}_{ij}$:
%%%%%%%%%%%
\beq
\int d^D x \sqrt{-\hat{g}} \left[ \frac{M_{(D)}^{D-2}}{2} \hat{R} - \frac{1}{2} {\cal G}_{ij} \hat{g}^{\mu\nu} \hat{\nabla}_\mu \phi^i \hat{\nabla}_\nu \phi^j - \hat{V} \right] ,
\label{actionmulti2}
\eeq
with
%%%%%%%%%%%%
\beq
{\cal G}_{ij} = \frac{M_{(D)}^{D-2}}{2f} \delta_{ij} + \frac{(D-1)}{(D-2)} \frac{M_{(D)}^{D-2}}{f^2} f_{,i} f_{,j} .
\label{calG}
\eeq
Note that since scalar fields have dimensions $[ \phi ] \sim (mass)^{(D-2)/2}$, then $[ f_{,i} ] \sim [ \partial_\phi f ] \sim (mass)^{(D-2)/2}  \sim [ \phi ]$.

A necessary condition for the existence of some conformal transformation that would bring the field-space metric into the desired form, ${\cal G}_{ij} \rightarrow \tilde{\cal G}_{ij} = \delta_{ij}$, is if the Riemann tensor constructed from the metric vanishes identically, $\tilde{\cal R}^i_{\>\>jkl} = 0$. \cite{Weinberg} Naturally the number of nontrivial components of the Riemann tensor grows rapidly with increasing $N$. We will therefore consider the Ricci curvature scalar constructed from the field-space metric, $\tilde{\cal R} = \tilde{\cal G}^{ij} \tilde{R}^k_{\>\>ikj}$. In general there could exist some metric, $\tilde{\cal G}_{ij}$, for which $\tilde{\cal R} = 0$ even though not all components of $\tilde{\cal R}^i_{\>\>jkl} = 0$; that is, $\tilde{\cal R}$ could vanish because of cancellations among various non-zero terms within the full Riemann tensor. But the converse is not true: there is no way in which $\tilde{\cal R} \neq 0$ if $\tilde{\cal R}^i_{\>\>jkl} = 0$. In other words, $\tilde{\cal R} \neq 0$ if and only if $\tilde{\cal R}^i_{\>\>jkl} \neq 0$. For our purposes --- to demonstrate that no conformal transformation exists that could bring $\tilde{\cal G}_{ij} = \delta_{ij}$ --- it is therefore sufficient to demonstrate that $\tilde{\cal R} \neq 0$.

The Ricci curvature scalar corresponding to the metric ${\cal G}_{ij}$ in Eq. (\ref{calG}) is rather complicated, involving many factors of $f$, $f_{,i}$, and $f_{,ij} = \partial^2 f / \partial \phi^i \partial \phi^j$. Our concern is whether the target field-space is conformally flat, a condition that is itself conformally invariant. Hence we may reduce some of the clutter by making a conformal transformation in field space, rescaling the metric as
%%%%%%%%%
\beq
{\cal G}_{ij} \rightarrow \tilde{\cal G}_{ij} = \frac{2f}{ M_{(D)}^{D-2}} {\cal G}_{ij}  = \delta_{ij} + \frac{2 (D - 1)}{(D - 2)} \frac{1}{f} f_{,i} f_{,j} .
\label{rescaleGij}
\eeq
For $N > 1$, the curvature scalar corresponding to $\tilde{\cal G}_{ij}$ takes the form
%%%%%%%%%%%%
\beq
\begin{split}
\tilde{\cal R}_{(N)} = \frac{2 (D - 1)}{L (\phi^i)} & \left[ (D - 2) A^{ijkl} \left( f f_{,ij} f_{,kl} - f_{,i} f_{,j} f_{,kl} \right)   \right. \\
&\quad\quad \left. + 2(D - 1) B^{ijklmn} \left( f_{,i} f_{,j} f_{,kl} f_{,mn} \right) \right] 
\end{split}
\label{RN}
\eeq
where
%%%%%%%%%%%%%%
\beq
L (\phi^i) = \left[ (D -2 ) f + 2 (D - 1) \sum_i f_{,i}^2 \right]^2 
\label{L}
\eeq
and
%%%%%%%%%%%%%
\beq
\begin{split}
A^{ijkl} &\equiv \left[ \delta^{ij} \delta^{kl} - \delta^{ik} \delta^{jl} \right] , \\
B^{ijklmn} &\equiv \left[ \delta^{ij} A^{klmn} + 2 \delta^{ik} A^{jmln} \right] .
\end{split}
\label{AB}
\eeq
In general, each of the terms involving $f$ and its derivatives depends on $\phi^i$. In order for there to exist a conformal transformation that could bring $\tilde{\cal G}_{ij} = \delta_{ij}$, we would need to have $\tilde{\cal R}^i_{\>\>jkl}$ (and hence $\tilde{\cal R}$) vanish everywhere in field space, independent of particular values of the fields $\phi^i$. Thus we can see why in general such a conformal transformation is unlikely to exist. However, the properties of $\tilde{\cal R}_{(N)}$ differ between $N = 2$ and $N > 2$, and they are worth considering separately.

Before proceeding, note that if only one among the $N$ fields has a non-minimal coupling, while the remaining $(N - 1)$ fields remain minimally coupled, then there will always exist some combination of conformal transformation and field rescalings such that in the new frame both gravitational and kinetic terms in the action assume canonical form. This result follows from the structure of $A^{ijkl}$: the terms in which all indices take the same value vanish, so that every remaining term includes derivatives of $f (\phi^i)$ along at least two directions in field-space. If $f (\phi^i)$ depends only on a single field, then every term within $\tilde{\cal R}_{(N)}$ will vanish (and one can show that the same holds for the full $\tilde{\cal R}^i_{\>\>jkl}$). Thus multifield models with only one non-minimally coupled field behave much like the usual single-field case. One important difference is that the scalar potential, $V (\phi^i)$, will acquire new interactions between the non-minimally coupled field and the minimally coupled fields, owing to the scaling $V \rightarrow \hat{V} = \Omega^{-D} V = [ 2 f (\phi) / M_{(D)}^{D-2} ]^{-D/(D - 2)} V (\phi^i)$. Now let us consider the case in which more than one field has a non-minimal coupling.

\subsection{$N=2$}

For $N = 2$, Eq. (\ref{RN}) for $\tilde{\cal R}$ simplifies considerably. The term involving $B^{ijklmn}$ vanishes identically, leaving
%%%%%%%%%%%
\beq
\tilde{\cal R}_{(2)} = \frac{2 (D - 1) (D - 2)}{L(\phi^i)} 
 \left[ 2f f_{,11} f_{,22} - f_{,1}^2 f_{,22} - f_{,2}^2 f_{,11} - 2 f_{,12} (f f_{,12} - f_{,1} f_{,2}) \right]  .
\label{RN2}
\eeq
Furthermore, for $N = 2$, we also find that $\tilde{\cal R}^i_{\>\>jkl} \propto \tilde{\cal R}_{(2)}$. In particular we have (no sum on repeated indices)
%%%%%%%%%
\beq
\begin{split}
\tilde{\cal R}^i_{\>\>iji} &= - \frac{(D-1)}{(D - 2)} \frac{ f_{,i} f_{,j} }{f} \> \tilde{\cal R}_{(2)} , \\
\tilde{\cal R}^i_{\>\>jij} &=  \left[ \frac{1}{2} + \frac{(D-1)}{(D-2)} \frac{f_{,j}^2}{f} \right] \tilde{\cal R}_{(2)}. 
\end{split}
\label{RiemannN2}
\eeq
The relationship $\tilde{\cal R}^i_{\>\>jkl} \propto \tilde{\cal R}$ will not generalize for $N > 2$.

Let us consider a typical form for $f (\phi^i)$ in the case $N = 2$. We may label the fields $\phi^1 = \phi$ and $\phi^2 = \chi$. Then the non-minimal couplings in the action typically take the form
%%%%%%%%%%
\beq
f(\phi, \chi) = \frac{1}{2} \left[ M_0^{D-2} + \xi_{\phi} \phi^2 + \xi_{\chi} \chi^2 \right] ,
\label{fN2}
\eeq
where the coupling strengths $\xi_{\phi}$ and $\xi_{\chi}$ need not coincide. For this typical form for $f$, the cross-term derivatives vanish, $f_{,12} = f_{,\phi \chi} =  0$, leaving
%%%%%%%%%
\beq
\begin{split}
L(\phi, \chi) \tilde{\cal R}_{(2)} &= 2 (D - 1) (D - 2) \left[ 2 f f_{, \phi  \phi} f_{, \chi \chi} - f_{,\phi}^2 f_{,\chi  \chi} - f_{,\chi}^2 f_{,\phi  \phi}   \right] \\
&= 2 (D - 1) (D- 2) \xi_{\phi} \xi_{\chi} M_0^{D-2} .
\end{split}
\label{RN2example}
\eeq
In a model with two scalar fields, each of them non-minimally coupled to the spacetime curvature as in Eq. (\ref{fN2}), we therefore see that no conformal transformation exists that can bring both the gravitational sector and the scalar fields' kinetic terms into canonical form. Even in the Einstein frame, in other words, the fields would not behave as they would in a genuine minimally coupled model.

On the other hand, in the case of $N = 2$, one could find a conformal transformation that would bring both the gravitational and kinetic terms into canonical form if $M_0 = 0$. In fact, for $N = 2$ and $M_0 = 0$, $\tilde{\cal R}^i_{\>\>jkm} = 0$ even if $\xi_{\phi} \neq \xi_{\chi}$, a relationship that does not generalize to models with $N > 2$. Examples include Jordan-Brans-Dicke gravity with two scalar fields, or induced-gravity models in which one or both of the fields has a non-zero vacuum expectation value, $v_i$, leading to $M_{(D)}^{D-2} = \sum_i \xi_i v_i^2$ below the symmetry-breaking scale.

\subsection{$N > 2$}

For arbitrary non-minimal coupling with $N > 2$, the $B^{ijklmn}$ term in Eq. (\ref{RN}) for $\tilde{\cal R}_{(N)}$ does not vanish. The $B^{ijklmn}$ term typically introduces terms in $\tilde{\cal R}_{(N)}$ that depend on the fields $\phi^i$ and not just their couplings, $\xi_i$. Generically, therefore, models with more than two non-minimally coupled scalar fields do not admit any conformal transformation that could bring $\tilde{\cal G}_{ij} = \delta_{ij}$.

Consider, for example, the $N > 2$ generaliztion of the typical non-minimal couplings of Eq. (\ref{fN2}), namely
%%%%%%%%
\beq
f (\phi^i) = \frac{1}{2} \left[ M_0^{D-2} + \sum_{i = 1}^N \xi_i (\phi^i)^2 \right] .
\label{fN}
\eeq
For this family of models, $f_{,i} = \xi_i \phi^i$ (no sum) and $f_{,ij} = \xi_i \delta_{ij}$, and the Ricci scalar becomes
%%%%%%%%%%%%
\beq
\begin{split}
\tilde{\cal R}_{(N)} = \frac{(D - 1)}{ L (\phi^i)} &\left[ (D - 2) M_0^{D-2} \sum_{ijkl} \left( \delta^{ij} \delta^{kl} - \frac{1}{N^2} \delta^{ik} \right) \xi_i \xi_k \right. \\
&\quad \left. + \sum_i \left( D - 2 + 4 (D - 1) \xi_i \right) \right. \\
&\quad\quad \left. \times  \sum_{jklm} \left[ \delta^{jm} \delta^{kl} - \frac{1}{N^2} \left( \delta^{jk} + \delta^{ij} + \delta^{ik} - 2 \delta^{ij} \delta^{ik} \right) \right] \xi_i \xi_j \xi_k (\phi^i )^2 \right] .
\end{split}
\label{RNmany}
\eeq
One might be tempted to search for particular combinations of the coupling constants $\xi_i$ that would yield exact cancellations and hence make $\tilde{\cal R}_{(N)} = 0$. However, for $N > 2$, it is no longer the case that $\tilde{\cal R}^i_{\>\>jkl} \propto \tilde{\cal R}_{(N)}$. The full Riemann tensor contains components such as
%%%%%%%%%%
\beq
\tilde{\cal R}^i_{\>\>ijk} = 2 (D - 1)^2  \frac{f_{,i}^2 f_{,j} f_{,k}}{f L} \left( f_{,jj} - f_{,kk} \right) 
\label{Riijk}
\eeq
for $i \neq j \neq k$ (no sum on $i$). The requirement that these components vanish is more stringent than requiring $\tilde{\cal R}_{(N)} = 0$ alone. For the family of models with $f (\phi^i)$ as in Eq. (\ref{fN}), terms like Eq. (\ref{Riijk}) will only vanish if all coupling constants are equal to each other: $\xi_i = \xi$ for all $i$, that is, if there exists an $O(N)$ symmetry among the $N$ non-minimally coupled fields. Under that requirement, Eq. (\ref{RNmany}) for $\tilde{\cal R}_{(N)}$ simplifies even further:
%%%%%%%%%%%%
\beq
%\begin{split}
\tilde{\cal R}_{(N)} = \frac{(D - 1)(N-1)}{ L (\phi^i)}  \left[ (D - 2) N  \xi^2 M_0^{D-2}  
 + \left[ D - 2 + 4 (D - 1)\xi \right] (N - 2) \xi^3 \sum_i (\phi^i )^2 \right] .
\label{RNxi}
%\end{split}
\eeq
For $N > 2$, $\tilde{\cal R}_{(N)} \neq 0$ even for models in which $M_0 = 0$. Thus the Riemann tensor does not vanish, and no conformal transformation exists that could make $\tilde{\cal G}_{ij} = \delta_{ij}$.

Although one cannot bring the action in the transformed frame into canonical form for arbitrary values of the fields, the Einstein-frame action will approach canonical form in the low-energy limit. For $f (\phi^i)$ as in Eq. (\ref{fN}), with $M_0 \neq 0$, we have
%%%%%%%%%%%
\beq
f (\phi^i ) \rightarrow \frac{1}{2} M_0^{D-2} + {\cal O} ( \xi_i (\phi^i )^2 / M_0^{D-2} ).
\label{fNlowenergy}
\eeq
From Eqs. (\ref{actionmulti2}) and (\ref{calG}), the kinetic terms in the Einstein frame would then become
%%%%%%%%%%%%%
\beq
- \frac{1}{2} \left( \frac{ M_{(D)} }{M_0} \right)^{D-2} \delta_{ij} \> \hat{g}^{\mu\nu} \hat{\nabla}_\mu \phi^i \hat{\nabla}_\nu \phi^j - \frac{2 (D - 1)}{(D -2 )} \left( \frac{ M_{(D)} }{M_0^2} \right)^{D-2} \xi_i^2 \left( \phi^i \hat{\nabla} \phi^i \right)^2 + ....
\label{kineticlowenergy}
\eeq
Eq. (\ref{kineticlowenergy}) generalizes a result found in \cite{Burgess,Hertzberg} for the specific case of ``Higgs inflation" in $D = 4$. In models for which $M_0 \sim M_{(D)}$, we would then recover the canonical kinetic terms plus corrections suppressed by $\xi_i^2 / M_{(D)}^{D-2}$. In the regime in which $\phi^i \ll M_{(D)}^{(D-2)/2} / \xi_i$, the effective action would then behave close to canonical form for minimally coupled fields, plus $\xi_i$-dependent corrections. Yet inflation in models like ``Higgs inflation" occurs for values $\phi^i \gg M_{(D)}^{(D-2)/2} / \sqrt{\xi_i}$, with $\xi_i = \xi \sim 10^4$ \cite{ShapBezrukov}. Thus in the regime of interest, we are far from the limit in which Eq. (\ref{kineticlowenergy}) applies, and Eq. (\ref{RNxi}) implies that no combination of conformal transformation and field rescalings can restore canonical kinetic couplings for all of the fields.

\section{Conclusions}

Scalar fields with non-minimal couplings are difficult to avoid. Such terms arise from a variety of model-building efforts, as well as from more formal requirements of renormalization. Of course, such models are tightly constrained by solar-system tests of gravitation as well as big bang nucleosynthesis \cite{Will,DamourBBN}. At some time in the history of our observable universe, in other words, the non-minimally coupled fields must have stopped varying appreciably, producing $f (\phi^i ) \rightarrow (16 \pi G_D)^{-1} \simeq {\rm constant}$. In this way, the Jordan frame must have evolved smoothly to an effectively Einsteinian one.

On the other hand, scalar fields with non-minimal couplings likely dominate the dynamics at very high energies or very early times, such as during early-universe inflation. To understand physics in these regimes, one cannot avoid the differences between the Jordan frame and the Einstein frame. In general, when more than two non-minimally coupled scalar fields are involved, we are not free to transform to a frame in which both gravitation and the fields' kinetic terms assume canonical form.  

We may come close if either the fields, $\phi^i$, or their couplings, $\xi_i$, behave in specific ways. We already noted in Eq. (\ref{kineticlowenergy}) that there are low-energy limits in which the transformed action relaxes toward canonical form, up to corrections that scale as $\xi_i^2 (\phi^i)^2 / M_{(D)}^{(D-2)}$. Even at higher energies, there might exist a particular point in field-space, $\phi_0^i$, that would make $\tilde{\cal R}^i_{\>\>jkl}$ vanish \cite{DamourNordvedt92,Burgess,Hertzberg}, even though $\tilde{\cal R}^i_{\>\>jkl}$ were non-zero in most regions of field-space; then for particular applications, in which the special point $\phi_0^i$ were relevant to the question of interest, one could move to an effectively canonical action. Likewise, if all but one of the non-minimally coupled fields became effectively frozen or varied slowly in the Jordan frame then one could transform to a frame in which the action approximated the case of a single non-minimally coupled field with $(N - 1)$ minimally coupled fields. However, the existence of such a special location in field-space should be demonstrated and not assumed; and to do that, one should study the coupled dynamics of the $N$-field system in the Jordan frame. 

Meanwhile, if all the couplings $\xi_i$ but one were small (or became small under renormalization-group flow), then the non-zero terms in $\tilde{\cal R}_{(N)}$ and $\tilde{\cal R}^i_{\>\>jkl}$ --- which typically scale as products of the coupling constants, $\xi_i \xi_j$, $\xi_i \xi_j \xi_k$, and $\xi_i \xi_j \xi_k \xi_l$ --- could become arbitrarily small, as can be seen from Eq. (\ref{RNmany}). In that case one could again transform to a frame in which the gravitational and kinetic terms in the action assumed canonical form, up to corrections suppressed by $\xi_i^2 / M_{(D)}^{(D-2)}$. If all but one of the $\xi_i$ were small, these correction terms could remain negligible even for high-energy interactions, such as during an inflationary phase.

In this analysis we have focused on models in which the non-minimally coupled scalar fields have canonical kinetic terms in the Jordan frame. One could broaden the investigation by considering general scalar-tensor models by replacing, for example, the constant Brans-Dicke parameter, $\omega$ in Eq. (\ref{SJBD}), by $\omega (\phi^i)$. Such a move would introduce new terms involving $\omega$, $\omega_{,i}$, and $\omega_{,ij}$ into the expressions for $\tilde{G}_{ij}$, $\tilde{R}^i_{\>\>jkl}$, and $\tilde{\cal R}_{(N)}$, which, like the corresponding terms involving $f(\phi^i )$ and its derivatives, would depend on $\phi^i$. Unless one chose the form of $\omega (\phi^i)$ in a highly ad hoc manner, the new Riemann tensor would still contain components akin to Eq. (\ref{Riijk}) which vanish when all $\xi_i = \xi$, that is, for models with an $O(N)$ symmetry among the non-minimally coupled fields. Since such target field-spaces are not flat for $N > 2$, the addition of non-canonical kinetic terms in the Jordan frame would not, in general, produce the needed cancellations that might bring $\tilde{\cal G}_{ij} = \delta_{ij}$. Thus the conclusions of this investigation should hold for general scalar-tensor models as well. Or, put another way, the burden would be to find a particular (ad hoc) model in which $\tilde{\cal R}^i_{\>\>jkl}$ happened to vanish because of arranged cancellations between the $\omega (\phi^i)$ and $f (\phi^i)$ terms. Absent such a model, one may choose to work in any frame that is convenient for a given problem, accepting either a non-canonical gravitational sector, non-canonical kinetic terms, or both.

\acknowledgements{It is a pleasure to thank Bruce Bassett for helpful discussions, and Ilya Shapiro, Daniel Zenh\"{a}usern, and Javier Rubio for helpful comments on an earlier draft. This work was supported in part by the U.S. Department of Energy (DoE) under contract No. DE-FG02-05ER41360.}

\end{document}